\documentclass[preprint,showpacs,preprintnumbers,amsmath,amssymb]{revtex4}
\usepackage{graphicx}
\usepackage{dcolumn}
\usepackage{bm}
\usepackage{epsfig}
\begin{document}

\title{Liquid polymorphism and  density anomaly in a  three-dimensional associating lattice gas}

\author{Mauricio Girardi}

\email{mgirardi@fge.if.usp.br}

\affiliation{Universidade Federal de Pelotas - UNIPAMPA/Bag\'e, Rua Carlos Barbosa SN, CEP 96400-970, Bag\'e, RS, Brasil.}

\author{Aline L. Balladares}

\email{linelb@if.ufrgs.br}

\affiliation{Instituto de F\'{\i}sica, Universidade Federal do Rio Grande do Sul, CP
15051, CEP 9105-970, Porto Alegre, RS, Brasil and Departament 
de Fisica Fonamental,  Universitat de Barcelona}

\author{Vera B. Henriques}

\email{vhenriques@if.usp.br}

\affiliation{Instituto de F\'{\i}sica, Universidade de S\~ao Paulo,
Caixa Postal 66318, 05315970, S\~ao Paulo, SP, Brazil}

\author{Marcia C. Barbosa}

\email{marcia.barbosa@ufrgs.br}

\affiliation{Instituto de Física, Universidade Federal do Rio Grande do Sul, CP
15051, CEP 9105-970, Porto Alegre, RS, Brasil.}

\begin{abstract}
We investigate the phase diagram of a three-dimensional  associating gas $(ALG)$  model.
This model combines orientational ice-like interactions and  ``van der Waals''
that might be repulsive, representing, in this case, a penalty 
for distortion of hydrogen bonds. These interactions can be interpreted
as two competing distances  making the connection between this
model and continuous isotropic soft-core potentials. We present Monte Carlo
studies of the $ALG$ model showing the presence of two liquid phase, two
critical points and A density anomaly.
\end{abstract}
\maketitle

\section{Introduction}

Water is an anomalous substance in many respects.
Most liquids contract upon cooling. This is not the case of water, a liquid
where the specific volume at ambient pressure starts to increase 
when cooled below $T=4 ^oC$ \cite{Wa64}. Besides, in a certain
range of pressures, water also exhibits an anomalous increase of compressibility 
and specific heat upon cooling \cite{Pr87}-\cite{Ha84}. Far less 
known are its dynamics anomalies: while for most
materials diffusivity decreases with increasing pressure,
liquid water has an opposite behavior in a large region
of the phase diagram 
\cite{St99}-\cite{Ne02}.

It was proposed a few years ago
that these anomalies are related to a second critical
point between two liquid phases, a low density liquid
(LDL) and a high density liquid (HDL) \cite{Po92}. This critical point 
was discovered
by computer simulations. This work   suggests that the
critical point is located at the 
supercooled region beyond  the line of
homogeneous nucleation and thus cannot be experimentally measured. 
Inspite of  this limitation,  this hypothesis has been supported by indirect experimental 
results \cite{Mi98}\cite{angell}.

Water, however, is not 
an isolated case. There are other examples of  tetrahedrally
bonded molecular liquids, such as phosphorus \cite{Ka00,Mo03}
and amorphous  silica \cite{La00}, that also are good candidates for having
two liquid phases. Moreover, other materials such as liquid metals
\cite{Cu81} and graphite \cite{To97} also exhibit thermodynamic anomalies.
Unfortunately a coherent and general interpretation of 
the low density liquid and high density liquid phases is still missing.

What kind of potential would be appropriated for 
describing the tetrahedrally bonded 
molecular liquids,  capturing the presence of 
thermodynamic anomalies? Realistic simulations of water
\cite{St74}-\cite{Jo00} have achieved a good accuracy in
describing the thermodynamic and dynamic anomalies of
water. However, due to the high number of microscopic
details taken into account in these models, it becomes
difficult to discriminate what is essential to explain
the anomalies. On the other extreme, a number of    
isotropic models were proposed 
as  the simplest framework to understand the physics of
the liquid-liquid phase transition and liquid state anomalies.
From the desire of constructing a
simple two-body isotropic potential capable of describing
the complicated  behavior present in water-like molecules,
a number of models in which
single component systems of particles interact via
core-softened (CS) potentials  have been proposed.
They possess a
repulsive core that exhibits a region of
softening where the slope changes dramatically. This region can
be a shoulder or a ramp \cite{St98}-\cite{Ca05}. Unfortunately, these 
models, even when 
successful in showing density anomaly and two liquid phases, fail
in providing the connection between the isotropic effective
potential and the realistic potential of water. 

It  would, therefore, be desirable to have a theoretical framework
which retains the simplicity of the core-softened potentials but
accommodates the tetrahedral structure and the role
played by the hydrogen bonds  present in water.  A
 number of lattice models in which the
tetrahedral structure and the hydrogen bonds 
are present  have been studied \cite{Be72}-\cite{Pr05}. One of them, is the
three-dimensional model proposed by Roberts and Debenedetti \cite{Ro96} and 
further studied by Pretti and Buzano \cite{Pr04} defined  on the body centered cubic
lattice. According to their approach,  the energy between two bonded
molecules rises when a third
particle is introduced on A site neighbour to
the bond. Using a cluster mean-field approximation, they were able
to find the density anomaly and two liquid phases. 

In order to investigate the mixtures of water with other chemical 
species as that present in a number of biological and industrial processes, it
would be interesting to have a  simpler model capable  of capturing
the same essential features observed in 
water and also being able to bridge the gap between
the realistic models for water and the isotropic softened-core potentials.
 
Thus, in this paper we investigate a three-dimensional associating lattice-gas model
that can fulfill both requirements.  Our model system is  a lattice gas with  ice
variables \cite{Be33} which allows for a low 
density ordered structure.
 Competition between the filling
up of the lattice and the formation of an open four-bonded orientational
structure
 is naturally introduced in terms of the ice bonding variables, and
no \emph{ad
 hoc} introduction of density or bond strength variations is
needed. In that sense, our approach 
 bares some resemblance to that of 
continuous softened-core models \cite{Si98,Tr99,Tr02}. Using
this simple model we are able to find two liquid phases, two critical
points and the density anomaly.
The remainer of the paper goes as follows. In sec. II the 
model is introduced and the simulation details are given. Sec. III is devoted to the main results and conclusion ends this session.

\section{The Model}

We consider a body-centered cubic lattice with $V$ sites,
where each site can be either empty or filled by a water molecule.
 Associated to 
each site there are two 
kinds of variables: an occupational variables, $\sigma_{i}$, and an
orientational one, $\tau_{i}^{ij}$. For $n_{i}=0$ the $i$ site
is empty, and $n_{i}=1$ represents an occupied site. The 
orientational state of 
particle $i$ is defined by
the configuration of its  bonding and 
 non-bonding arms, as illustrated 
in Fig 1.
 Four of them
are the usual ice bonding arms with $\tau_{i}^{ij}=1$ 
 distributed in a tetrahedral arrangement,
 and four additional arms are taken as 
inert or non-bonding ($\tau_{i}^{ij}$=0). Therefore, each
molecule can be in one of two possible states  $A$ and $B$ are shown
illustrated in Fig. 1.
A potential energy $\varepsilon$ is associated to any pair of occupied
nearest-neighbor ($NN$) sites, mimicking the van der Waals potential.
Here, water molecules have four indistinguishable arms that can form
hydrogen-bondS (HB). An HB  is formed  when two arms of $NN$ molecules
are pointing to each other with $\tau_i^{ij}=1$. An energy
$\gamma$ is assigned to
each formed HB.

\begin{figure}
\begin{center}
\includegraphics[height=9cm,width=9cm]{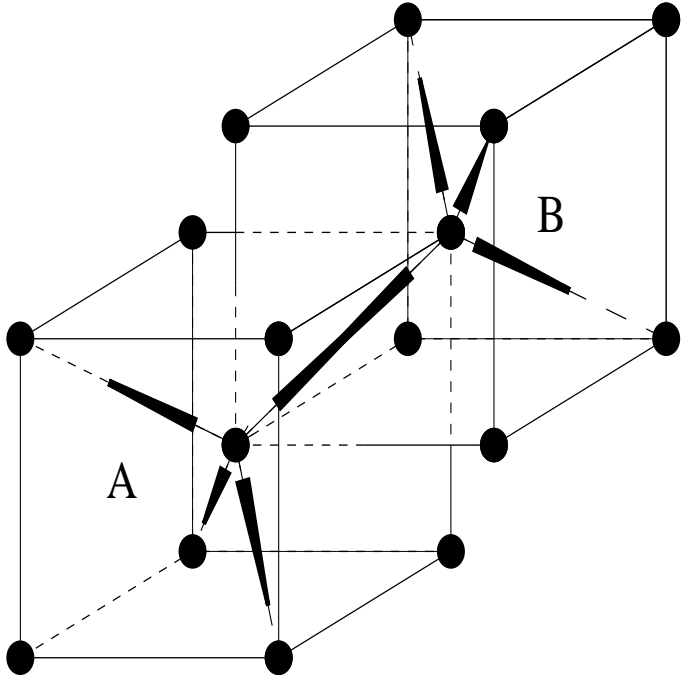}
\caption{The two possible states of the water molecules in the body centered
cubic lattice. $A$ and $B$ molecules are forming an hydrogen bond
since two of their arms are pointing each other.} 
\label{fig1} 
\end{center}
\end{figure}
In resume the total energy of the system is given by:
\begin{equation}
E=\sum_{(i,j)}n_{i}n_{j}\left(\varepsilon
+\gamma\tau_{i}^{ij}\tau_{j}^{ji}\right)\,\,.
\label{en}
\end{equation}
  The interaction parameters were chosen
to be $\varepsilon>0$ and $\gamma<0$, which implies in an energetic
penalty on neighbors that do not form $HBs$. 
From this
condition results the presence of two liquid phases
and the density anomaly.

The ground state of the system can be inferred by simply inspecting
the equation \ref{en}, and taking account an external chemical potential
$\mu$. At zero temperature, the grand potential per volume
 is $\Omega=e+\mu\rho$,
where $\rho$ is the water density and $e=E/V$. At very low values of the 
chemical potential, the lattice is empty and the system 
is in the gas phase. As the chemical potential increases, at
$\mu=-2(\varepsilon+\gamma)$,
a gas phase with $\rho=0$ and a low density liquid ($LDL$) with $\rho=1/2$
coexist. In this case, each molecule in the $LDL$ phase has four occupied
$NN$ sites, forming four $HBs$, and the energy per site is 
$e=\varepsilon+\gamma$. As the chemical potential increases
even further a competition between the chemical potential that favors
filling up the lattice and the HB penalty that favors molecules
with only four $NN$ sites appears. At
$\mu=-6\varepsilon-2\gamma$, the $LDL$ phase coexists
with a high density liquid ($HDL$) with $\rho=1$. In the $HDL$, each
molecule has eight $NN$ occupied sites, but forms only four $HBs$. The
other four non-bonded molecules are repealed, which can be viewed
as an effective weakening of the hydrogen bonds due to distortions
of the electronic orbitals of the bonded molecules. The energy per
molecule is then $e=4\varepsilon+2\gamma$.

Our model may be interpreted in terms of some sort of average soft-core potential
for large hydrogen-bond energies. The low density  phase implies an average interparticle
distance \( \overline{d_{LD}}=\rho _{LD}^{-1/3}=2^{1/3} \), whereas for
the high density  phase we have \( \overline{d_{HD}}=\rho _{HD}^{-1/3}=1 \). The corresponding
energies per pair of particles are  $e_p^{LDL}=\varepsilon+\gamma$ 
and $e_p^{HDL}=\varepsilon+\gamma/2$
respectively.  Figure 2 illustrates this effective pair
potential for the case of $\gamma/\varepsilon=-2$.  The hard core
is offered by the lattice, since two particles cannot occupy the same site.

\begin{figure}
\begin{center}
\includegraphics[height=9cm,width=9cm]{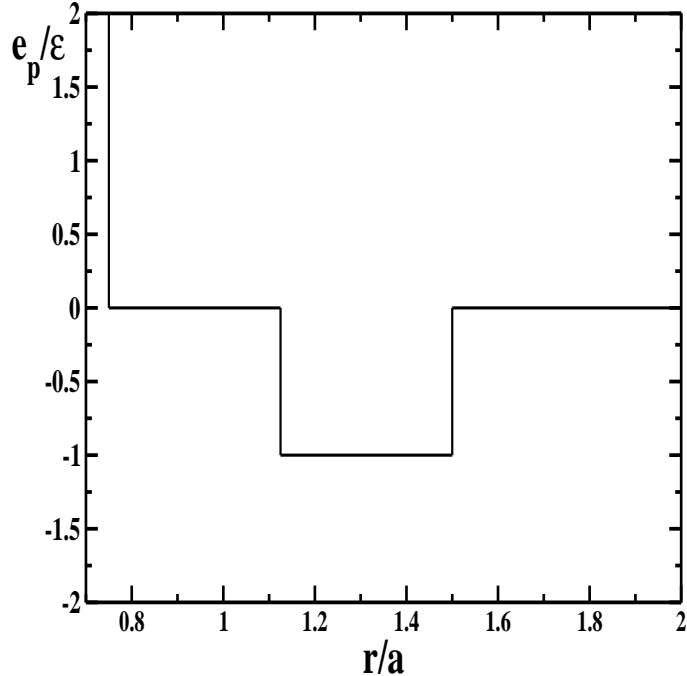}
\caption{Effective potential vs inter-particle distance for $\gamma/\varepsilon=-2$.The corresponding
energies per pair of particles are  $e_p^{LDL}=\varepsilon+\gamma$ 
and $e_p^{HDL}=\varepsilon+\gamma/2$ for $LDL$ and $HDL$ respectively.} 
\label{fig2} 
\end{center}
\end{figure}

The system pressure $P$
can be calculated from the grand potential since $\Omega=-P$. At
the gas-$LDL$ coexistence, $P=0$ and at $LDL$-$HDL$ coexistence point,
$P=2\varepsilon$.

The model properties for finite temperatures were obtained through Monte Carlo
simulations in the grand-canonical ensemble (chemical potential
and temperature were kept constant).
The total number of molecules is  allowed to change in time by means
of the Metropolis algorithm, where in one time unit (1 Monte Carlo
step) we test all lattice sites in order to insert or exclude one
water molecule. The insertion and exclusion transition rates are written
as $w(insertion)=exp(-\Delta\phi)$ and $w(exclusion)=1$ if $\Delta\phi>0$,
and $w(insertion)=1$ and $w(exclusion)=exp(+\Delta\phi)$ if $\Delta\phi<0$.
Here $\Delta\phi=\beta\left(e_{molecule}-\mu\right)-\ln\left(2\right)$,
where $e_{molecule}$ is the energy of the included (or excluded) molecule,
and the factor $\ln\left(2\right)$ guarantees the detailed balance. 

The simulations were carried out for lattices with linear size $L=10$,
and the interaction parameters were set to $\gamma/\varepsilon=-2$.
Runs were of the order of $10^{4}$ Monte Carlo steps. Some test 
runs were done for L=20, showing no relevantchange in the critical 
temperatures. A detailed
study of the model properties and the full phase diagrams was undertaken for
an L=10 lattice.

In order to obtain the pressure-temperature phase diagram of the model, the 
pressure was calculated from the simulation data. By numerical integration of
the Gibbs-Duhem relation, $SdT-VdP+Nd\mu=0$ at fixed temperature,
we obtain $P(\rho,T)$, using the condition that $P=0$ at $\rho=0$.
Since the model presents two first-order phase transitions (from gas
to $LDL$, and from $LDL$ to $HDL$ phases), the curves $\rho$ versus $\mu$
have two discontinuities and hysteresis loops. The hystereses were
observed when the simulations were started at different initial conditions
for a given chemical potential around the transition point.

\section{Results and Conclusions}

The model properties for finite temperatures that were obtained through simulations at constant temperature and chemical potential go as follows.
For sufficiently low values of the chemical potential and at low temperatures,
all attempts to insert molecules are frustrated, and the total density
$\rho$ remains equal to zero. By increasing the value of $\mu$,
the  molecules begin to enter in the system, increasing  $\rho$  and leading to
two first-order transitions, one between the gas and the $LDL$ phases
and another between the $LDL$ and $HDL$ phases.
The dependence between  $\rho$ and the reduced
chemical potential $\bar{\mu}=\mu/\varepsilon$ for some 
  temperatures is illustrated in Fig. 3a. Similarly the 
number of hydrogen bonds per site is illustrated in Fig. 3b. 
The transition between one hydrogen bond per site to 
two hydrogen   bonds per site occurs at the $LDL$-$HDL$ phase transiton.
The coexistence of the gas and $LDL$ phases and the $LDL$ and the $HDL$ phases
were then obtained  from this data.

In order to confirm the loci of the coexistence lines and the critical
points, the histograms of the densities were 
collected during a simulation run. The histograms for four different
temperatures and chemical potentials are shown in Fig. 4 for
illustration. Near
a first-order phase transition, mainly inside the metastable region,
the histogram is double-peaked, and the system density fluctuates
around two characteristic values. One can obtain the coexistence lines
by finding the chemical potential in which both peaks have the same
height (Figs. 4a and 4c). As the temperature approaches the critical
value, the peaks converge to a single one, and a homogeneous phase
appears (Figs. 4b and 4d).

\begin{figure}[htb]
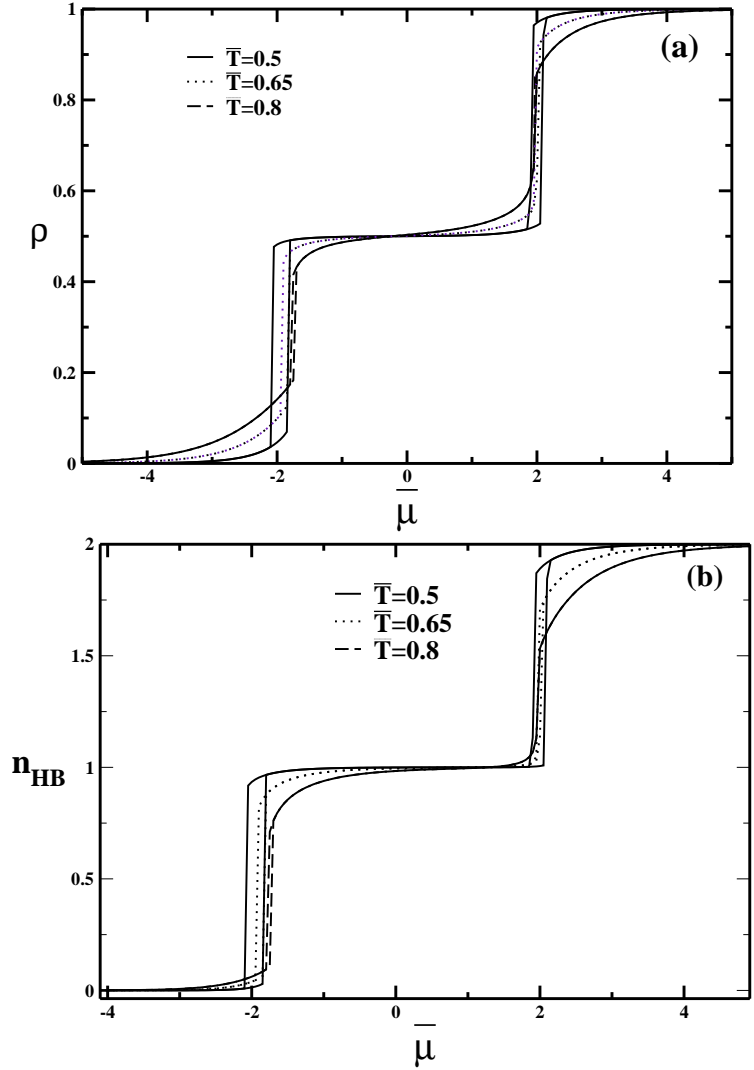

\center
\includegraphics[clip=true,scale=0.4]{fig3a.eps}
\hspace{1cm}\includegraphics[clip,scale=0.4]{fig3b.eps}
\caption{(a) Density isotherms vs. reduced chemical potential for different temperatures. (b) Number of bonds per site vs. reduced chemical potential for different temperatures.
$\rho$ is given in units of lattice space and the temperature is in units
of $k_B$.} 
\label{fig3} 
 \end{figure}

\begin{figure}
\begin{center}
\includegraphics[height=9cm,width=9cm]{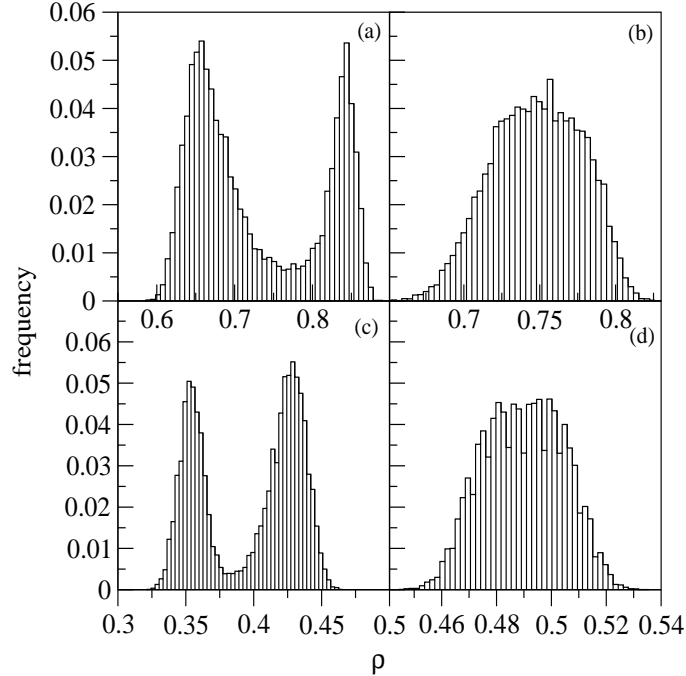}
\caption{Histograms of the total density $\rho$. (a) the coexistence between
$LDL$ and $HDL$ phases at $T=0.8$ and (b) an homogeneous phase at $T=1.0$
near the $LDL-HDL$ critical temperature. (c) the coexistence between $gas$ and
$LDL$ phases at $T=1.2$ and (d) an homogeneous phase at $T=1.4$, near
the $gas-LDL$ critical temperature.} 
\label{fig4} 
\end{center}
\end{figure}

In Fig. 5 we exhibit the reduced  pressure, $\bar{P}=P/\varepsilon$, versus density  isotherms. The gas-$LDL$ and  $LDL$-$HDL$ first-order phase
transitions are evidenced by the presence of 
 plateaus in the $\bar{P}.vs.\rho$ curves
at low reduced temperatures, $\bar{T}=T/k_B$. 

\begin{figure}
\begin{center}
\includegraphics[height=9cm,width=9cm]{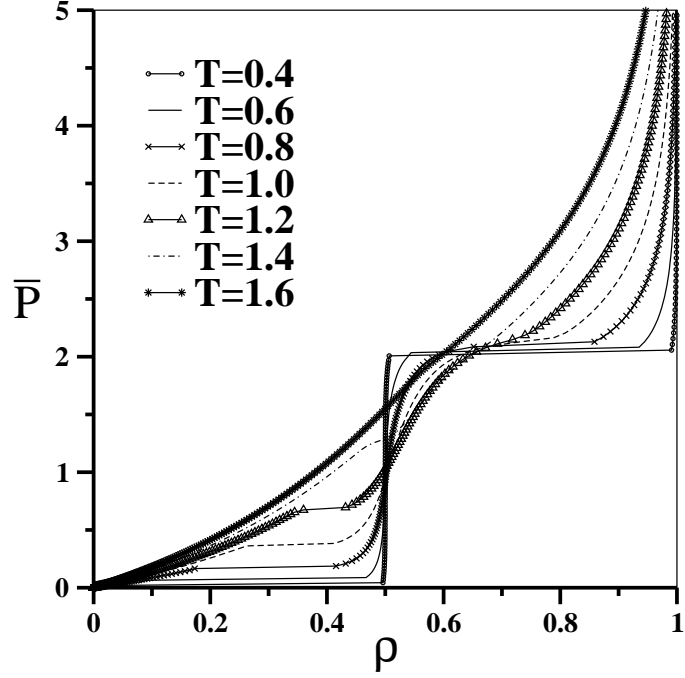}
\caption{Reduced pressure  as a function of the total density $\rho$ for some
values of $\bar{T}$.} 
\label{fig5} 
\end{center}
\end{figure}

The plot of  density versus reduced temperature  at 
constant pressures
 shows that an inversion of the behavior of density as
a function of temperature takes place at intermediate pressures, in the $LDL$
phase. At smaller pressures, density decreases with
temperature, whereas at higher pressures,  density increases
with temperature. This yields a temperature of maximum
density for a fixed pressure, $TMD$, in the higher range of
pressures, which we illustrate  in Fig. 6.

\begin{figure}
\begin{center}
\includegraphics[height=9cm,width=9cm]{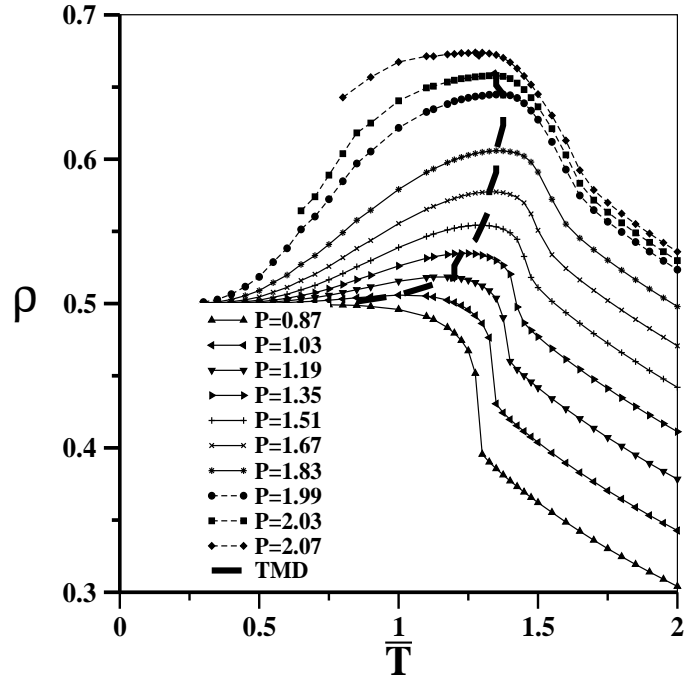}
\caption{ Total density as a function of reduced temperature at constant values
of the reduced pressure. The maximum in the curves give the temperature of
maximum density  for a given pressure. } 
\label{fig6} 
\end{center}
\end{figure}

The  pressure-temperature phase diagram is illustrated
in Fig. 7. The gas, $LDL$ and $HDL$ phases are shown together
with the two coexistence lines, the  two  critical
points and the line of temperature of maximum density $(TMD)$ as 
a function of pressure. Density  versus reduced  
temperature illustrating the 
two coexistence regions and the two critical points are shown in Fig. 8.

\begin{figure}
\begin{center}
\includegraphics[height=9cm,width=9cm]{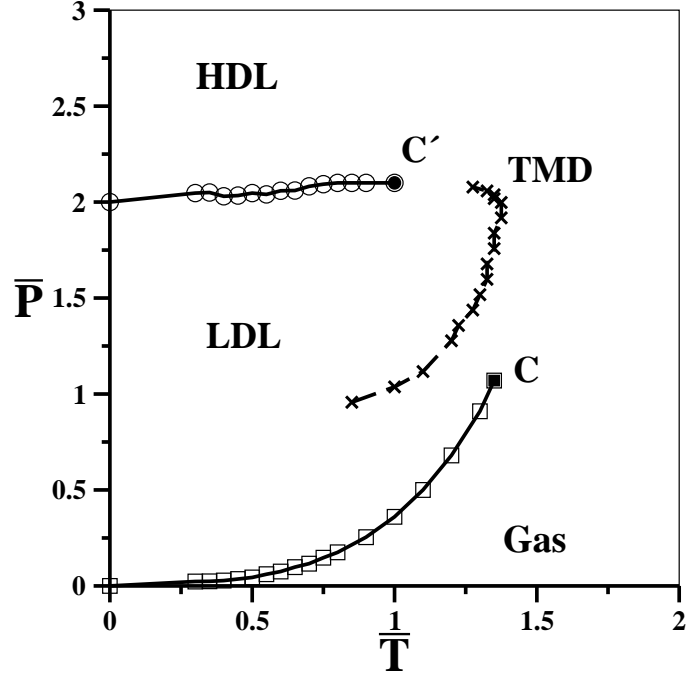}
\caption{ Reduced pressure versus reduced temperature phase diagram. The gas-$LDL$ and $LDL$-$HDL$ coexistence lines, the two critical points and the TMD line are 
shown. } 
\label{fig7} 
\end{center}
\end{figure}

\begin{figure}
\begin{center}
\includegraphics[height=9cm,width=9cm]{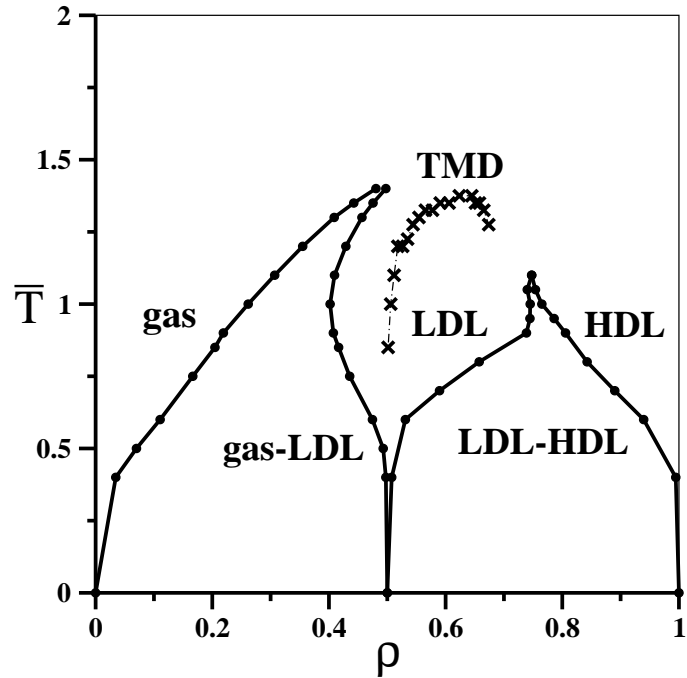}
\caption{ Density as versus reduced  temperature illustrating the 
two coexistence regions,  the two critical points and the TMD line.  } 
\label{fig8} 
\end{center}
\end{figure}

We  have shown that it is possible to incorporate some of the microscopic properties of true water molecules into a very simple minimal model that still
contains some of the ingredients of real water without having 
its whole complexity. 

The model  includes orientational
and occupational variables, and guarantees the local distribution of hydrogens
on molecular bonds, without the need of increasing the volume artificially or
introducing artificial orientational variables \cite{Bes94}. In spite of the
absence of an orientational order-disorder transition \cite{Na91}, the model
presents liquid-liquid coexistence, with slightly positive slope in the pressure-temperature
plane, accompanied by a line of maximum density  on the low density side, a
feature expected for real water. Besides, this study points out to the fact that
the presence of a density anomaly, with the thermal expansion coefficient \( \alpha <0 \), on the low temperature
side, and as a consequence, \( (\frac{\partial S}{\partial p})_{T}>0 \), \emph{does
not} imply a negative slope of the liquid-liquid line, contrasting with the
results for most studies of metastable liquid-liquid coexistence in models for
water, which suggest a transition line with negative gradient \cite{Sc00}.

The presence of both a density anomaly and two liquid phases in our model begs
the question of which features of this potential are responsible for such behaviour.
Averaged over orientational degrees of freedom, our model can be seen as some
kind of shoulder potential, with the liquid-liquid coexistence line being present
only for a repulsive \"{ }van der Waals\"{ } potential. The same was indeed
observed for continuous step pair potentials \cite{Fr01}\cite{Fr02}, for which,
however, the density anomaly is absent. On the other hand, a density anomaly
has been observed in a number of shoulder-like 
lattice models in which the major
ingredient is the competition between two scales \cite{Ba04,Ol05,Ol06}. This
feature is present in our case. Therefore it seems that the 
competition between two scales is the major ingredient that
warrantIES the presence of the density anomaly. If, in addition, the model
has an  attractive interaction, two liquid phases and two critical
points emerge.

\vspace*{1.25cm}

\noindent \textbf{\large Acknowledgments}{\large \par}

\vspace*{0.5cm} This work was supported by the Brazilian science agencies CNPq,
FINEP, Fapesp and Fapergs.

\end{document}